\documentclass[conference]{IEEEtran}
\IEEEoverridecommandlockouts
\usepackage{cite}
\usepackage{amsmath,amssymb,amsfonts}
\usepackage{algorithmic}
\usepackage{graphicx}
\usepackage{textcomp}
\usepackage{xcolor}
\usepackage{listings}
\usepackage{graphics}
\usepackage{amssymb}
\usepackage{graphicx}
\usepackage{color}
\usepackage{caption}
\usepackage{subcaption}
\usepackage{multicol}
\usepackage{multirow}
\usepackage{url}
\usepackage{booktabs}
\definecolor{darkgreen}{rgb}{0, 0.5, 0}
\usepackage{booktabs}        
\usepackage{soul}

\usepackage{listings}

\def\BibTeX{{\rm B\kern-.05em{\sc i\kern-.025em b}\kern-.08em
    T\kern-.1667em\lower.7ex\hbox{E}\kern-.125emX}}
\begin{document}

\newcommand{\apdnote}[1]{{\color{blue}[ #1 -- APD ]}}

\author{\IEEEauthorblockN{Adrian P. Dieguez\IEEEauthorrefmark{1}, Min Choi\IEEEauthorrefmark{2}, Mahmut Okyay\IEEEauthorrefmark{2},  Mauro Del Ben\IEEEauthorrefmark{1}, Bryan M. Wong\IEEEauthorrefmark{2}, and Khaled Z. Ibrahim\IEEEauthorrefmark{1}} 
\IEEEauthorblockA{ \IEEEauthorrefmark{1}
Lawrence Berkeley National Laboratory, CA,  USA \\
\textit{Corresponding Email: aperezdieguez@lbl.gov}} 
\IEEEauthorblockA{\IEEEauthorrefmark{2}
University of California-Riverside, Riverside, CA, USA\\
} 
}



\title {\huge Cost-Effective Methodology for Complex Tuning Searches in HPC: Navigating Interdependencies and Dimensionality}

\maketitle
\pagestyle{plain}
\thispagestyle{plain}

\begin{abstract}

Tuning searches are pivotal in High-Performance Computing (HPC), addressing complex optimization challenges in computational applications. The complexity arises not only from finely tuning parameters within routines but also potential interdependencies among them, rendering traditional optimization methods inefficient. Instead of scrutinizing interdependencies among parameters and routines, practitioners often face the dilemma of conducting independent tuning searches for each routine, thereby overlooking interdependence, or pursuing a more resource-intensive joint search for all routines. This decision is driven by the consideration that some interdependence analysis and high-dimensional decomposition techniques in literature may be prohibitively expensive in HPC tuning searches.  Our methodology adapts and refines these methods to ensure computational feasibility while maximizing performance gains in real-world scenarios. Our methodology leverages a cost-effective interdependence analysis to decide whether to merge several tuning searches into a joint search or conduct orthogonal searches. Tested on synthetic functions with varying levels of parameter interdependence, our methodology efficiently explore the search space. In comparison to Bayesian-optimization-based full independent or fully joint searches, our methodology suggested an optimized breakdown of independent and merged searches that led to final configurations up to 8\% more accurate, reducing the search time by up to 95\%. When applied to GPU-offloaded Real-Time Time-Dependent Density Functional Theory (RT-TDDFT), an application in computational materials science that challenges modern HPC autotuners, our methodology achieved an effective tuning search. Its adaptability and efficiency extend beyond RT-TDDFT, making it valuable for related applications in HPC.

\end{abstract}

\section{Introduction}
\label{intro}

Significant attention is focused on HPC landscapes in the exascale era, where numerous scientific applications exhibit scalability challenges. To fully harness the capabilities of exascale supercomputers, these applications need to tune a wide range of performance parameters. One exemplary case is Real-Time Time-Dependent Density Functional Theory (RT-TDDFT) applications, which are extensively used in chemistry, physics, and materials science domains, making up to 70\% of the computing time used at NERSC \cite{nersc}.  Autotuning has traditionally accomplished this task by either empirical searches or analytical models. However, these methods are becoming infeasible due to the complexity of large search spaces, and programmers started adopting mathematical optimization methods to explore the search space in an intelligent manner. Bayesian optimization (\texttt{BO}) \cite{BO20} has gained popularity as it explores only certain promising regions of the space, delivering high-performing configurations within a short number of evaluations. In practice, \texttt{BO} does not tackle problems with high dimensionality (number of parameters). While this upper limit depends on many factors, authors in \cite{moriconi2020highdimensional} emphasized the significance of 20. Nevertheless, tuning searches that involve a high number of parameters are becoming common in exascale, and the challenge lies not just in the precise tuning of parameters but also in navigating potential relationships among them. Relying on heuristics becomes challenging in scenarios where there are performance interdependencies among parameters. Analyzing these interdependencies among parameters and routines comes at a high cost \cite{pmlr-v37-kandasamy15}, leading practitioners to choose between two extreme options to avoid such analysis: running a single joint search for all kernels or opting for independent tuning searches. This dilemma results in either missing tuning opportunities or incurring additional costs for high-dimensional tuning, which are impractical in certain scenarios.

This work aims at addressing tuning searches involving a substantial number of performance parameters and routines to be optimized, with a specific emphasis on RT-TDDFT applications designed for GPU-based supercomputers. Distributed computing together with GPU kernels introduce a vast parameter space that necessitates optimization. The computational pattern in \textit{RT-TDDFT applications} is used in several frameworks and algorithms\cite{tddft_nanodroplet}. Therefore, tuning lessons learned from one framework can be applied to others. As such, we have developed a bottom-up approach comprising multiple steps to handle complex tuning searches across many routines within an application.  The approach strives to be high-level and adaptable to other disciplines with similar tuning patterns to broaden its value to the HPC community. Additionally, as the trend of offloading RT-TDDFT workloads to GPUs gains momentum in recent times and there is a limited number of RT-TDDFT applications characterized by an extensive search space, we demonstrate the effectiveness of our methodology by tuning a GPU version of a RT-TDDFT application for two different material inputs. To underscore the versatility of our  tuning methodology, we also target a set of synthetic functions with different interdependence levels, demonstrating the applicability of our approach across a diverse array of optimization problems and outperforming full-independent and fully-joint extreme cases.

Our main contribution in this paper is to introduce a methodology for tackling complex \texttt{BO}-based tuning searches across different routines within an application, which surpass capabilities of current HPC autotuners, that reduces the required observations while offering potential applicability to other domain-related tuning searches. We novelly apply a bottom-up analysis to infer orthogonality or interdependence between tuning parameters, based on sensitivity to individual variations, that saves multiple application evaluations compared to a traditional orthogonal analysis. This analysis creates a Directed Acyclic Graph (DAG) partition problem that decides whether to merge several routine searches into a joint higher-dimensional search or keep them independent. Employing our methodology, we focus on the tuning of a representative RT-TDDFT application, accelerated through offloading to GPUs, when scaling across multiple resources on US Department of Energy (DOE) supercomputer, such as Perlmutter.

\section{Related Work}

The domain of autotuning is very rich in algorithms and methods for fast tuning searches, such as random search. However, random search, along with other approaches such as grid search, has been demonstrated to be not as accurate as Bayesian optimization (\texttt{BO}) when finding optimal parameters in massive search spaces \cite{liashchynskyi2019}. Significant effort has been made in recent years to apply \texttt{BO} to complex scenarios, leading to three main different strategies \cite{9555522} related to search dimensionality. Some approaches \cite{10.5555/2540128.2540383,10.5555/3495724.3495855} exploit an embedded strategy where the algorithm optimizes a low-dimensional subspace to identify the next candidate and then is projected back to the original dimensions to evaluate the objective. However, these projections can create distortions when evaluating the objective function. 
Another option \cite{10.5555/3172077.3172179} is to perform the search over $d$ out of $D$ dimensions in every iteration, filling the remaining dimensions with random values, which leads, in general, to slower convergence rate. Other approaches \cite{pmlr-v37-kandasamy15} are based on decomposing a complex search as the sum of independent low-dimensional functions. However, the independence analysis leads to a substantial number of observations, which can incur significant costs in an HPC context. 
This work inverts this idea by merging low-dimensional functions that show interdependence, while controlling the resulting dimensionality to ensure optimal performance. By doing this,  current \texttt{BO}-based frameworks for HPC can be re-used to this end. Additionally, our interdependence study reduces the required observations for the analysis. The existing strategies previously explained lack a dedicated tuning framework for HPC. Conversely, \texttt{BO}-based HPC autotuners do not incorporate these strategies. There are several autotuners based on \texttt{BO} in the HPC community, such as ytopt~\cite{wu2023ytopt} and DeepHyper \cite{Dorier_2022}. We have selected GPTune~\cite{gptune} due to its advantageous features, such as crash recovery, effective search space constraints, and the capability for transfer learning, which has previously proven to be beneficial for RT-TDDFT computations\cite{PMBS}.


\section{Background}

This work aims to address the challenge of tuning searches for specific routines within an application, where the total number of parameters amounts to 20 or more, for instance, an HPC application supporting the RT-TDDFT computational pattern. To address the resource-intensive nature of collecting configuration evaluations for an HPC application, we  illustrate our methodology using synthetic functions to support wide exploration of approaches while mitigating the impact of computational requirements of the full application. Additionally, we use \texttt{BO} to run the resulting tuning searches.

\subsection{Bayesian Optimization}
Bayesian optimization (\texttt{BO}) \cite{BO20} is an optimization-search application for finding the optimal parameters that optimize a given objective function. A surrogate model captures the behavior of the objective function. It is initially trained using a small set of initial configurations, randomly sampled from the search space. Subsequently, an acquisition function guides the selection of the next configuration to be evaluated. It balances the exploration of not-evaluated regions with the exploitation of the promising ones. The suggested configuration is evaluated, re-training the surrogate model for improved accuracy. The acquisition suggestion and model updating are interactively repeated until a stopping criterion is met.  The cost of tuning applications 
can be substantial. 
However, tuning overhead is usually amortized across multiple or interactive executions. 
\texttt{BO} reduces this cost 
by guiding the search  through a decision-making process. 
Constraints on the search reduce exploration regions and avoid infeasible solutions. Nevertheless, aggressive constraints could confine the search within local minima and create additional overhead during the search. This adds an extra layer of complexity that requires careful consideration. As previously stated, \texttt{BO} is not well suited for solving high-dimensional tuning problems \cite{moriconi2020highdimensional}. In this work, we used the open-source GPTune framework \cite{gptune} to execute Bayesian optimization searches.

\subsection{ Density Functional Theory}
Density Functional Theory (DFT) is a powerful numerical approach that 
enables accurate and efficient electronic structure and material properties predictions, impacting %
physics and chemistry. Its strength lies in the description of many-particle systems through the
electron density distribution \cite{KS}. Density is characterized in relation to single-electron wavefunctions arrays, which are specified by state-bands and k-points -- representing the crystal momentum vector in solids \cite{Bloch}. Wavefunctions are expressed as a sum of plane-wave basis functions, with their determination relying on G-vectors. The Hamiltonian is a functional dependency on the density, resulting in a system of nonlinear differential equations solved iteratively in a self-consistent manner.
Overall, the dominant operations in this workload are reductions, dense linear algebra operations, and Fast Fourier Transforms.  Real-Time Time-Dependent DFT (RT-TDDFT) computationally simulates optical properties and electronic excitations in material and chemical systems \cite{tddft_nanodroplet}. In general, an RT-TDDFT simulation starts from an initial DFT ground state calculation. In contrast to ground-state DFT, RT-TDDFT calculates the time-dependent wavefunction under the influence of an external perturbation.

\subsection{Synthetic Objective Functions}

\begin{figure*}[t!]

  \begin{equation*} 
F(x_0,...,x_{19}) = \underbrace{\sum_{i=0}^3 (x_i-x_{i+1})^2 + \sum_{i=0}^4 A_i  }_{Group 1: \, x_0 \cdots x_4} + \underbrace{\sum_{k=5}^{8} (x_k-x_{k+1})^4 + \sum_{k=5}^{9} A_j}_{Group 2: \, x_{5} \cdots x_{9}} + \underbrace{\sum_{u=10}^{14} \boxed{ \; \; \; \;\; \; \; \; }}_{Group 3: \, x_{10} \cdots x_{14}} + \underbrace{\sum_{v=15}^{19} 1/x_v + \epsilon}_{Group 4: \, x_{15} \cdots x_{19} }
  \end{equation*}

    \captionsetup{font=small}
    \caption{Synthetic 20-dim function body where $A_i=10\cdot cos(2\pi\cdot(x_i -1))+\epsilon$, with $\epsilon$ representing random noise. The template box outlined with \textit{Group 3} is replaced with a set of implementations to create different synthetic functions.}
    \label{equation}
    \vspace{-2mm}
\end{figure*}

Synthetic functions have been broadly employed in tuning literature to test autotuning proposals \cite{Ginsbourger2010,GPTuneCrowd}. The availability of HPC applications offering more than 20 tunable parameters is currently limited and their tuning process can incur significant computational expenses. Therefore, this work employs a set of five 20-dimensional synthetic functions, allowing for a comprehensive benchmark without incurring  substantial computational costs. The structure of these functions is depicted in Figure \ref{equation}, where the empty box is redefined for each synthetic case, and each $x_i$ variable can take any real between -50 and 50. Introducing random noise serves to augment the modeling complexity, aligning with the inherent unpredictability encountered in HPC applications. It is also a common practice in HPC to have distinct kernels or code regions executing independently, offering the opportunity for separate optimization. However, as previously explained, discerning whether the tuning parameters influencing these regions are interdependent is not always clear, requiring of complex, and sometimes inconclusive, analyses. To simulate this characteristic behavior, the equation depicted in the figure decomposes the overall objective into four distinct \textit{groups} that represent different kernels or regions within an application that contribute to the overall objective value. Specifically, each \textit{group} encompasses its own set of variables, representing the visible performance parameters that each routine may tune:  $\{x_0 \cdots x_4\}$ for Group 1; $\{x_5 \cdots x_9\}$ for Group 2; $\{x_{10} \cdots x_{14}\}$ for Group 3; and $\{x_{15} \cdots x_{19}\}$ for Group 4. Notably, Group 3 is uniquely implemented for each synthetic function. The definitions of Group 3 for each synthetic case are detailed in Table \ref{group3_defs}, contributing to formulate the overall synthetic expression illustrated in Figure \ref{equation}.

\begin{table}[h]
  \centering
\renewcommand{\arraystretch}{1.5}

  \begin{tabular}{|c|c|c|}
    \hline
    \textbf{Name} & \textbf{Group 4's influence} & \textbf{Group 3 Formula} \\
    \hline
    Case 1 & Very Low & $\sum_{u=10}^{14} x_u + \sum_{v=15}^{19} cos(2\pi\cdot x_v) +\epsilon$ \\
    \hline
    Case 2 & Low & $\sum_{u=10}^{14} x_u^2 + \sum_{v=15}^{19} x_v+\epsilon$  \\
    \hline
    Case 3 & Medium & $\sum_{u=10}^{14} x_u^2 + \sum_{v=15}^{19}x_v^2+\epsilon$ \\
    \hline
    Case 4 & High & $\sum_{u=10,v=15}^{14,19} (x_ux_{v}^4)^2 +\epsilon$ \\
    \hline
    Case 5 & Extremely High & $\sum_{u=10,v=15}^{14,19} (x_ux_{v}^8)^2+\epsilon$ \\
    \hline
  \end{tabular}
  \captionsetup{font=small}
  \caption{Group 3 definition for each synthetic case, with $\epsilon$ random noise, and the corresponding influence from Group 4 variables.}
  \label{group3_defs}
\end{table}

Breaking down each \textit{group}'s equation in Figure \ref{equation} reveals discernible interdependence patterns between variables within a group (routine), dictated by multiplication or power functions, while also orthogonality among them, created by additive operations. For example, variables within Group 1 exhibit a dependent relationship among them. In addition to relationships \textit{within groups}, the interdependence can also happen \textit{across groups}: Group 3 presents a unique case where, in addition to its own variables($x_{10} \cdots x_{14}$), variables from Group 4 ($x_{15} \cdots x_{19}$) also play a role in computing its objective value. This mirrors tuning scenarios where performance parameters from other routines impact the performance of a given routine, establishing an interdependence between routines. It is noteworthy that Group 4 variables can either be orthogonal (Cases 1, 2, 3) or non-orthogonal (Cases 4, 5) with Group 3 variables, while always impacting the Group 3 objective value. Table \ref{group3_defs} delineates five distinct cases in which Group 4 variables exert varying degrees of influence on the Group 3 value. For instance, in Case 1, Group 4 variables are encapsulated within a cosine function, resulting in a minimal contribution to the overall Group 3 value --indicating a low-influence scenario. On the other hand, Group 4 variables in Case 3 contribute in a manner equivalent to Group 3 variables, signifying a medium-influence scenario. Meanwhile, Group 4 variables in Cases 4 and 5 contribute exponentially, establishing a high-influence scenario. Finally, a \texttt{log()} transformation is applied to the absolute value of each group's result. It becomes clear that higher exponents within the equation increases the equation sensitivity to these variables. The subsequent sections will shed light on the efficacy of our methodology in navigating such synthetic functions.

\section{Methodology for Tackling Complex Tuning Searches}
 
 The following subsections present the methodology that we followed to tackle the challenge of complex tuning searches with Bayesian optimization by reducing the number of required observations compared to other approaches:

\begin{enumerate}
    \item Constrain the search with a domain expert and define the maximum cost of the tuning search.
    \item Perform statistical analyses to get insights about tuning parameters and runtime.
    \item Search for kernels or routines that may exhibit interdependence between them: A sensitivity analysis can help with finding interdependence.
    \item Merge dependent searches and drop parameters: We limit to 10 dimensions per search.
    \item If the same kernel appears in different regions, and its parameter values must be the same across all regions, prioritize the kernel with highest impact.
 
\end{enumerate}


Our methodology follows a bottom-up approach for tuning $t$ routines within an application. Instead of adopting $t$ full-independent searches, ignoring the potential interplay between routines, or pursuing a single fully joint search, increasing the cost and complexity, our methodology analyzes the interdependence between them. It intelligently merges searches when appropriate, resulting in an optimized set of searches to be conducted. This approach effectively reduces the amount of required evaluations while still achieving favorable results. The methodology consists of two distinct phases. Firstly, our independence analysis helps with finding interdependence between all parameters and routines. It assigns an score of influence to each parameter on every routine. Subsequently, the second phase uses this information to merge and run the corresponding searches based on the influence score and the chosen search method, which is \texttt{BO} in our case. 
It is noteworthy that tuning searches can be very complex due to numerous factors, and we are not striving for a one-size-fits-all solution. However, given the common patterns across different RT-TDDFT algorithms, we believe this guideline holds potential for applicability within many of them.

\subsection{Domain Knowledge and Complexity}

Domain knowledge allows the identification of the most influential range of parameter values and limits the search space with valid solutions. Tuning must be done in an agnostic manner considering the optimization as a blackbox while using the expert's inputs to avoid wasting computational resources. While synthetic functions can offer a cost-effective approach, high-dimensional searches can incur significant costs; hence, HPC practitioners usually set a predetermined computing budget to constrain the search complexity.

\subsection{Insights about parameters}
\label{insights}

Sensitivity analysis \cite{saltelli} is used twice in our methodology and involves studying the impact of a parameter on the runtime. By quantifying the \textit{sensitivity} of the application to the parameters' variations, we can identify which parameters significantly influence the execution. First, we establish one configuration as a baseline, and then test $V$ different variations individually on each parameter, calculating the average runtime variability per parameter as $\frac{1}{V}\times \sum^V_{i=1} |(time_{baseline} - time_i)/time_{baseline}|$. Section \ref{orthogonality} will also use sensitivity analysis for inferring independent components.

Additionally, statistical analysis and feature importance provide valuable insights into search complexity. Parameters lacking importance can be removed, while correlated ones might be grouped in a search. It is essential to be cautious when interpreting results made on top of data samples; thus, we have used the one-in-ten rule \cite{harrell2013regression}, a general guideline that suggests that building regression models would need at least 10 observations for each independent variable.  Feature importance complements correlation, capturing intricate relationships and overcoming skewed distributions. On the one hand, feature importance analysis highlights the contributions of each feature to the accuracy of machine-learning models \cite{feature}. Parameters with the highest score should be conserved in the search for better accuracy.  On the other hand, a Pearson correlation analysis reveals linear relationships. While more intricate analyses like partial correlation exist, they require larger samples. \\

\textit{Synthetic Functions - Insights}. With respect to the sensitivity analysis, a baseline configuration was randomly selected, and subsequently, 100 individual variations were systematically applied to each parameter. Each variation involved increasing the variable value by 10\% relative to the preceding iteration. When running this analysis on the 5 synthetic functions, it discerns a notable trend where variables affected by superior power exponent show higher sensitivity, as already explained. Pearson correlation aligns with expectations, revealing the absence of linear dependence between variables. Concurrently,  a feature importance analysis, leveraging Random Forest trees, was also conducted, which showed a uniform distribution of modeling importance across variables.

\begin{table}
    \centering
    \resizebox{6cm}{!}{%
    \begin{tabular}{|l|l|l|l|l|l|}
        \hline
       \textbf{Feature}&\textbf{Case 1} & \textbf{Case 2} & \textbf{Case 3} &\textbf{Case 4} &\textbf{Case 5} \\
        
       \hline
        $x_{10}$ & 108\% &  82\% &   67\% &  13\% &  1.36\% \\
        $x_{11}$ & 72\% &  91\% &    84\% &  14\% &  1.46\% \\
       $ x_{12}$ & 100\% &  96\% &   83\% &  20\% &  0.01\% \\
       $ x_{13}$ & 104\% &  91\% &   87\% &  31\% &  0.01\% \\
        $x_{14}$ & 90\% &  92\% &    83\% &  26\% &  0.03\% \\
        \hline
        $x_{15}$& 1.99\% &  13\% &  46\% &  104\% &  120\% \\
        $x_{16}$ & 1.51\% &  10\% &  81\% &  107\% &  86\% \\
       $ x_{17}$ & 1.95\% &  4.83\%& 68\% &  114\% &  122\% \\
        $x_{18}$ & 1.77\% &  14\% &  85\% &  126\% &  77\% \\
       $ x_{19}$ & 1.61\% &  3.40\% &84\% &  121\% &  79\% \\
      
        \hline
    \end{tabular}%
 }
 \captionsetup{font=small}
\caption{Variability of Group 3 output for the 5 synthetic cases. Showing the top 10 sensitive variables, which always corresponded with Group 3 ($x_{10}\cdots x_{14}$) and Group 4 ($x_{15} \cdots x_{19}$) variables.}
\label{tab:sensitSynth}
 \vspace{-4mm}
\end{table}

\subsection{Inferring independent routines}
\label{orthogonality}

Conducting an orthogonality analysis for an HPC application can be resource-intensive, requiring numerous observations \cite{pmlr-v37-kandasamy15}. In this work, we aim to mitigate the observation burden while preserving efficiency to understand the interdependence between tuning routines. To do so, we novelly leverage sensitivity analysis to infer routine orthogonality by analyzing how their variations impact the runtimes. By studying the individual effect of each parameter on every routine baseline configuration, we significantly reduce the required observations. This balances a trade-off between efficiency and capturing more complex influences. It can be conceptualized as a partitioning problem on Directed Acyclic Grahps (DAGs), where vertices represent routines, and their edges denote how their parameters affect the runtime variability of routines. Although most edges (parameters) are expected to connect to their own vertex (own routine), instances where one vertex connects to others indicate a strong parameter-performance dependence, necessitating their joint search. To avoid weak performance impacts on other vertices or runtime fluctuations, we implement an edge-pruning mechanism based on a cut-off. 

External parameters influencing routine performance offer two potential approaches. One option is to tune affected parameters twice: first within their routine and then within the external routine impacted by their performance. This maintains two independent searches, but one has higher dimensionality. However, this approach isn't universally applicable. In situations where parameters must share the same value across the entire application, both routine searches must be merged into a single search to minimize joint runtime, thereby reducing the overall application runtime.\\

\textit{Synthetic Functions - Independent routines}. Performing additional sensitivity analyses for each \textit{group} on every synthetic case revealed distinct patterns.  Although there is a marginal variability (less than 1\%) produced by the noise introduced in the formula, there is no representative influence from external variables in Group 1, 2 and 4 outputs. However, the sensitivity analysis of Group 3 reveals a different trend. Table \ref{tab:sensitSynth} depicts the variability of Group 3's objective value based on the fluctuations of every variable. Results align with expectations, demonstrating the capability of the sensitivity analysis to detect interdependence between routines. In Cases 1 and 2,  the primary source of variability comes from its own variables ($x_{10}..x_{14}$). In Case 3, variability is equally influenced by variables from both Group 3 and 4. Conversely, in Cases 4 and 5, most of the variability in Group 3's output is determined by variables from Group 4 ($x_{15}..x_{19}$). This indicates that the tuning of Group 1 and Group 2 can be conducted separately for all synthetic cases, while Group 3 and 4 show interdependence between them. Sensitivity analysis results align with the different grades of influence associated to the five synthetic cases, demonstrating its usability to detect tuning independence between routines at low cost.

\subsection{Establishing the ultimate set of tuning searches}

Once the interdependencies are detected, the next step involves determining whether to perform a merged search for the routines that show interdependence. This decision depends on the achieved score of influence, with a cut-off value determining when to keep them separate or to merge them. This cut-off value is influenced by various factors, including the nature of the application, the search mechanism employed, and the available computing budget. The larger dimensionality of a joint search, the more data is required for an effective modeling of the interdependencies. While Bayesian optimization stands out as a promising search mechanism in HPC, the training complexity of Gaussian Processes, used as surrogate models in \texttt{BO}, is $O(N^3)$, where $N$ denotes the number of datapoints (configuration evaluations). If there are not enough evaluations, the potential benefits of running a joint search may not outweigh the advantages of conducting two independent searches that navigate a reduced search space, which can be conducted in parallel. There is no a one-size-fits-all cut-off, it depends on the specific characteristics of the problem at hand. In scenarios where, after applying the cut-off to the DAG, the resulting set of searches suggests tuning over 10 parameters (dimensions) for a given routine, our methodology opts to choose the ten most influential variables (based on the data insights from Section \ref{insights}) for the search, while assigning default tuning values to the discarded variables. The choice of ten dimensions in our methodology is grounded in the feasibility of conducting outstanding \texttt{BO} searches within a manageable number of iterations for our target applications.\\ 

\begin{figure}[t!]
\centering 
\includegraphics[scale=0.35]{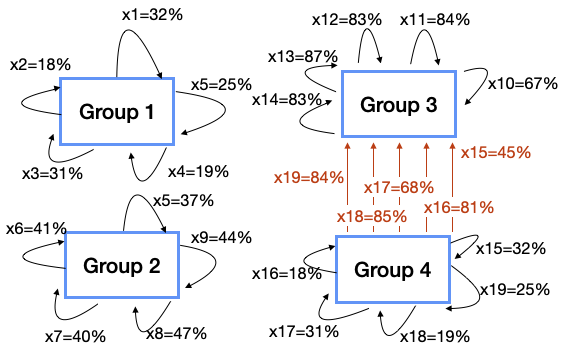}
\captionsetup{font=small}
\caption{The DAG diagram illustrates the interdependencies between variables and groups during the tuning process for synthetic Case 3 (medium influence), after applying a 25\% cut-off.}
\label{dag-synth2}
\end{figure}

\textit{Synthetic Functions -  Establishing the set of searches}.  Building upon prior analyses, we considered a 25\% cut-off to delineate interdependencies among variables and groups. Figure \ref{dag-synth2} illustrates the resulting DAG for the synthetic Case 3 (medium influence of Group 4 variables on Group 3). Consequently, a joint tuning search for Group 3 and Group 4 was established on synthetic Cases 3, 4 and 5, while independent searches were run for the other scenarios. We conducted the \texttt{BO} searches on the GPTune framework, starting the training with 5 random configurations. Several conversion criteria exist for stopping the \texttt{BO} search; we decided to complete a fixed number of evaluations by evaluating at least $10\times num\_parameters$ configurations during each \texttt{BO} search.

As mentioned earlier, practitioners typically opt for either full independent searches, denoted as \textit{G1,G2,G3,G4} here,  or a singular fully joint search, represented as \textit{G1+G2+G3+G4}. To showcase the efficacy of our methodology, we compare the recommended search composition against these two extreme approaches, where \textit{G1,G2, G3+G4} signifies independent searches on Group 1 and 2, and a merged search for Group 3 and 4. As Table \ref{BO-synth} illustrates, Bayesian optimization consistently outperformed Random Search across all scenarios. The marginal improvement of the 20-dim \texttt{BO} search (with $N=200$) over Random Search in some instances is attributed to its high dimensionality, leading to less efficient navigation of the space. Higher-dimensional \texttt{BO} searches necessitate more evaluations for better modeling of the extensive search space, while the $O(N^3)$ training complexity penalizes the search time. Inherent sequentiality made \texttt{BO} slower than parallelizable Random Search.  Instead, the \textit{G1,G2, G3+G4} approach, which runs three parallel searches of $N=\{50,50,100\}$, outperformed the fully independent search strategy, four \textit{N=50} searches in parallel, in Cases 4 and 5. In Case 3, both approaches found a similar minimum. In Cases 1 and 2, where interdependence between Groups 3 and 4 was weak, no significant accuracy improvement  was observed. In both strategies, the search space is significantly reduced compared to a 20-dimension joint search, facilitating efficient navigation in less evaluations. Although the achieved improvement may seem modest, the optimal values lie in the left tail of their runtime distributions, making their attainment very challenging. In summary, our methodology effectively suggested an optimized set of tuning searches, as evidenced by the results.

\begin{table}[h]
    \centering
      \resizebox{9cm}{!}{%
    \begin{tabular}{|c|c|c|c|c|c|c|c|c|}
        \hline
        Case &\multicolumn{2}{|c|}{Random Search} & \multicolumn{2}{c|}{G1+G2+G3+G4 BO}&\multicolumn{2}{|c|}{G1, G2, G3+G4 BO} & \multicolumn{2}{c|}{G1, G2, G3, G4 BO} \\
        \hline
        & Minima Found & Time & Minima Found & Time & Minima Found & Time & Minima Found & Time\\
        \hline
        Case 1&19.4&11.32&18.4&1468&15.3&79.11&\textbf{14.1}&\textbf{48.66}\\
        \hline
        Case 2&25.8&11.78&24.5&1580&18.1&196.52&\textbf{17.2}&\textbf{51.1}\\
        \hline
        Case 3&27.7&14.25&24.8&1760&\textbf{20.1}&\textbf{181.69}&20.1&51.3\\
        \hline
        Case 4&53.2&12.74&51.2&1598&\textbf{44.3}&\textbf{194.46}&45.4&52.24\\
        \hline
        Case 5&74.4&15.59&73.2&1501&\textbf{69.6}&\textbf{153.61}&75.4&53.27\\
        \hline
       
        \hline
    \end{tabular}%
    }
    \captionsetup{font=small}
    \caption{Minima found and corresponding search time 
    (seconds) for the 5 synthetic cases by testing different search strategies, averaging the results on 5 executions. Approaches suggested by our methodology are highlighted for each case, which are not necessarily the
best.}
    \label{BO-synth}
    \vspace{-4mm}
\end{table}


\section{The RT-TDDFT application}

\begin{figure}[t!]
\centering 
\includegraphics[scale=0.35]{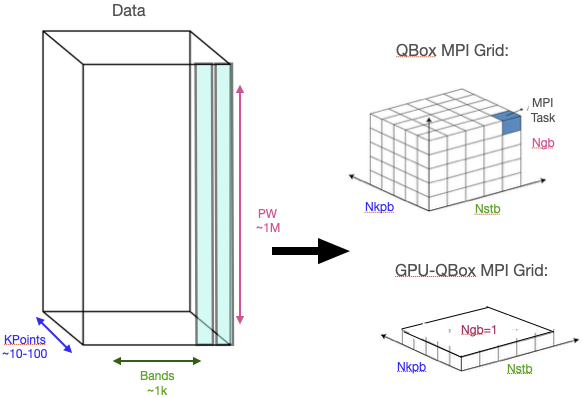}
\captionsetup{font=small}
\caption{Mapping of the wavefunction computation (left) into MPI tasks (right). The original CPU MPI partition is on the top, while the GPU partition is on the bottom. The spin dimension is 1 to enable a 3D representation. }
\label{mpi}
 \vspace{-6mm}
\end{figure}

Complex RT-TDDFT applications can be implemented by using augmented routines in the open-source QBox code \cite{qbox}.
This framework represents each wavefunction by a 4-dimensional, double-complex matrix, which is defined by spin, k-point, state-bands, and plane-wave (G-vector) dimensions, as Figure \ref{mpi} depicts. The size of each dimension is determined by the physical system. The parallelization in QBox involves distributing the wavefunction computation among MPI tasks, which creates a four-dimensional MPI grid of $nspb\times nkpb \times nstb \times ngb$ dimension, as Figure \ref{mpi} illustrates with the \textit{QBox MPI Grid}. The computation consists of several inner loops that compute an overall reduction and strongly hinges on parallelizing operations (relying on rank parallelism, using MPI communications). When profiling different executions of QBox with MPI parallelism, our results reveal that around 40-50\% of the runtime is attributed to communication primitives. Notably, most of this overhead is incurred during a matrix transpose\&padding step when calculating 3D-FFTs among $ngb$ MPI tasks, which is a recurrent component of the application. Specifically, this component is the Slater Determinant's energy potential computation for all bands of each k-point.  Figure \ref{qbox-pattern} summarizes the pseudo-code of the dominant computational pattern, where this Slater Determinant computation occurs in Lines 5-22. The Slater Determinant computation is integral to the QBox implementation, thereby influencing not only the RT-TDDFT modification but any DFT execution utilizing this framework.

\lstset{
    xleftmargin=1.5em,
    language=python,
    numbers=left,
    basicstyle=\fontsize{5pt}{6pt}\ttfamily, 
    breaklines=true,
    keywordstyle=\color{blue}\ttfamily,
    moredelim=[is][\color{orange}]{@O}{@},
    moredelim=[is][\color{red}]{@R}{@},
    moredelim=[is][\color{blue}]{@B}{@},
    commentstyle=\color{darkgreen}\textit, 
    escapeinside={\%*}{*)}
}
\begin{figure}[t!]
\centering
\begin{lstlisting}
for all rtiterations:
    while !SCF_converged:
        for all spins_loc:
            for all kpoints_loc:
                for all bands_loc:
                    @O#Group 1:@
                    #memcpy(HtoD)
                    map_vector_to_zvec() # cuVec2Zvec
                    fft_backward_z() # cuFFT-3D
                    bwd_transpose() #cuZcopy
                    fft_backward_xy() #cuFFT-3D
                    @B#Group 2:@
                    pairwise_multiplication() #cuPairwise
                    @R#Group 3:@
                    fft_forward_xy() #cuFFT-3D,cuDscal
                    fwd_transpose() #cuZcopy
                    fft_forward_z() #cuFFT-3D,cuDscal
                    map_zvec_to_vector() #cuZvec2Vec
                    #memcpy(DtoH)
                    
                    daxpy()
                    ... accumulations and MPI reductions...
\end{lstlisting}
\captionsetup{font=small}
\caption{Pseudo-code for the QBox-based RT-TDDFT that summarizes its dominant computational pattern.}
\label{qbox-pattern}
 \vspace{-4mm}
\end{figure}

\subsection{GPU Offloading of RT-TDDFT Kernels}

The GPU version of the RT-TDDFT creates a high-dimensional tuning space by offloading several kernels to the GPU. Offloading efforts are towards the recurrent Slater Determinant computation. To better align with the GPU programming model, the distributed-memory computation for the 3D-FFT is replaced with a shared-memory 3D-FFT approach. As the 2D- and 1D-FFT kernels from Figure \ref{qbox-pattern} are now packed into a single 3D-FFT invocation of the GPU cuFFT library, this presents an opportunity for fine-tuning this routine. Additionally, the shared-memory code refactoring substitutes the $ngb$ ranks working on transposing the matrix with a single-rank GPU computation. Therefore, the MPI $ngb$ parameter is set to $ngb=1$ in the GPU version (see \textit{GPU-QBox Grid} in Figure \ref{mpi}), disrupting the optimal balance among previous MPI parameters and necessitating the exploration of a new value for the MPI partition. There are other five CUDA kernels in the Slater Determinant offloading: \texttt{cuVec2Zvec} (moving data from one domain structure to the other), \texttt{cuZcopy} (used during the matrix transpose\&padding operations), \texttt{cuDscal} (essential for coefficient scaling in cuFFT), \texttt{cuPairwise} (pairwise multiplication), and \texttt{cuZvec2Vec}. The detailed description of each kernel is not essential in this work, the focus must be directed towards the tuning parameters associated with each kernel. Each kernel can be tuned with three different parameters, which are related to the loop \textit{unrolling} factor, \textit{threadblock size}, and number of active \textit{threadblocks per Streaming Multiprocessor} (SM). Furthermore, as the local computation of the different bands (Line 5) is inherently independent, data are packed in \textit{batches} for a single kernel invocation (Line 5),  providing additional avenue for finding its optimal size. Furthermore, the computation of several iterations can be overlapped through different \textit{CUDA\_streams}, whose optimal value must be found.

As a point of reference, the \texttt{cuFFT} kernel accounts for the 61.4\% of the GPU computing time (excluding overhead of memory transfer), followed by \texttt{cuZcopy} at 14.2\%, \texttt{cuVec2Zvec} at 12.4\%, \texttt{cuPairwise} at 4.9\%, \texttt{cuDscal} at 4.2\%, and \texttt{cuZvec2Vec} at 2.9\%, all of which use default tuning values for their kernels. In summary, the MPI grid partition will determine the number of local spins, k-points, and bands to be computed by each MPI rank (Lines 3-5), while the GPU tuning parameters will affect the performance of the defined GPU kernels and memory transactions (Lines 7-19). 

The tuning of this application is grounded in three main reasons. Firstly, the performance of MPI is constrained by the vendor runtime, which lies beyond the scope of application developers. 
Secondly, as we offload newly dominant kernels to the GPU, the search space's dimensionality will exacerbate and performance parameters may interplay due to cache memory interdependencies, which cannot be captured by traditional heuristics. Finally, this application is executed multiple times, thereby, the long-term impact of accelerating all involved kernels results in significant savings of computing hours.

\section{Defining the RT-TDDFT Search Space}

Table \ref{parameters} contains the 20 performance parameters associated with our RT-TDDFT GPU implementation. These parameters are MPI-related ($nstb,nkpb,nspb$) and GPU-related (\textit{u-}unroll, \textit{tb-}threadblock size, and \textit{tb\_sm-}threadblocks per SM for each kernel, together with $nbatches$ and $nstreams$).  Depending on the total number of MPI ranks, given as  $N_{nspb}\times N_{nstb} \times N_{nkpb}$, the range of possible values varies and, consequently, the search space. The range of values for the GPU parameters is architecture dependent (see Section \ref{setup}), leading to a total search space of possible configurations of $41,943,040\times N_{nstb} \times N_{nkpb} \times N_{nspb}$. Nevertheless, not all configurations are valid. Different \texttt{BO} frameworks handle this differently: some frameworks may need to explore configurations to identify their validity, while others might set constraints initially to only explore valid configurations, adding an extra layer of complexity to the search. In this search, $nstb\cdot nkpb \cdot nspb$ must be less than the total number of allocated cores for the execution, and each GPU kernel configuration must satisfy $tb\cdot tb\_sm$ to be lower than the maximum number of active threads per SM allowed by the GPU architecture. 

Similar to the synthetic functions, the dominant pattern in RT-TDDFT  exhibits a potential tuning division into distinct \textit{groups} that seemingly have independent tuning characteristics (see pseudo-code in Figure \ref{qbox-pattern}), suggesting a separate GPU tuning: \textit{Group 1} (\texttt{ZCOPY,VEC} related parameters), \textit{Group 2} (\texttt{PAIR}), and \textit{Group 3} (\texttt{ZCOPY,DSCAL,ZVEC}). To analyze the insights of this search space, we analyze the parameter influence on four regions: the Slater Determinant runtime (Lines 5-20 of Figure \ref{qbox-pattern}), and the GPU kernels grouped into Group 1, Group 2, and Group 3 in Figure \ref{qbox-pattern}. 

\begingroup
\linespread{0.7}
\begin{table}[t!]
     \small
    \centering
    \begin{tabular}{p{4.68cm}p{3cm}}
        \hline
        Parameter & Configurations \\
        \hline
        \texttt{nstb}, \texttt{nkpb}, \texttt{nspb} & $N_{nstb} \times N_{nkpb} \times N_{nspb}$ \\
        \texttt{u\_DSCAL, tb\_DSCAL, tb\_sm\_DSCAL} & $4 \times 32 \times 32$ \\
        \texttt{u\_PAIR}, \texttt{tb\_PAIR}, \texttt{tb\_sm\_PAIR} & $4 \times 32 \times 32$ \\
        \texttt{u\_ZCOPY, tb\_ZCOPY, tb\_sm\_ZCOPY} & $4 \times 32 \times 32$ \\
        \texttt{u\_VEC}, \texttt{tb\_VEC}, \texttt{tb\_sm\_VEC} & $4 \times 32 \times 32$ \\
        \texttt{u\_ZVEC}, \texttt{tb\_ZVEC}, \texttt{tb\_sm\_ZVEC} & $4 \times 32 \times 32$ \\
        \texttt{nstreams}, \texttt{nbatches} & $32 \times 32$ \\
        \hline
        Total Configurations & $41.943.040\times$\\ &$N_{nstb} \times N_{nkpb} \times N_{nspb}$ \\
        \hline
    \end{tabular}
       \captionsetup{font=small}
    \caption{RT-TDDFT tuning parameters and amount of possible configurations.}
    \label{parameters}
   \vspace{-3mm}
\end{table}

\endgroup

\section{Computational Setup}
\label{setup}

Results have been measured on Perlmutter at NERSC. We have specifically targeted the "GPU nodes", equipped with a single AMD EPYC 7763 CPU featuring 64 cores (x2 hyperthreading) and 256 GB of DDR4 at 204.8 GB/s. Each node is currently composed of four NVIDIA Ampere A100 GPUs with a PCI-e 4.0 GPU-CPU connection, and nodes are interconnected with Cray Slingshot 11 interconnect fabric. To prevent interferences among MPI tasks, we have restricted each GPU to a single task, resulting in 4 MPI tasks per node. The remaining cores are assigned for shared-memory OpenMP computations. The A100 GPU enables up to 32 active threadblocks per SM and up to 32 warps per threadblock, which constrain the range of possible values for previously defined GPU parameters. Finally, we chose two common physical systems for our case studies; one is a molecular 0D system, namely a magnesium porphyrin molecule comprised of one magnesium, 20 carbon, 4 nitrogen, and 12 hydrogen atoms (Case Study 1), and the other is a periodic 2D slab of 4$\times$4 hexagonal boron-nitride with 32 atoms per supercell (Case Study 2). Case Study 1 is composed of 1 spin, 1 k-point, 64 bands, and an FFT size of 3 million double complex elements; whereas Case Study 2 has 1 spin, 36 k-points, 64 bands, and an FFT size of 620k double complex elements.

\section{Experimental Results: Applying the Methodology to RT-TDDFT}

In this section, we apply our methodology to the introduced GPU-offloaded RT-TDDFT application for the \textit{Case Study 1} and \textit{Case Study 2} inputs,  described at Section \ref{setup}. To optimize computational resources during the tuning search, a single iteration of the outer loop (\textit{rtiterations} in the pseudocode shown in Figure \ref{qbox-pattern}) is executed.\\

\vspace{-3mm}
\textit{Domain Knowledge and Complexity}

In the tuning case of RT-TDDFT, an application expert ascertained the optimal range of values for the application-related parameters based on the given physical system, while a GPU expert constrained the exploration of GPU-specific parameters, such as the unrolling factor or workload per kernel. In our study, domain experts confine the search space by incorporating realistic configurations to reduce the search of $41.943.040\times N_{nstb} \times N_{nkpb} \times N_{nspb}$ configurations. It should be observed that this approach differs from substituting the tuning process with expert heuristics. Regarding computing budget,  we restricted the search to a maximum of 10 computing nodes and, as conversion criteria, we decided to complete $10\times num\_parameters$ configurations during each \texttt{BO} search.\\

\vspace{-3mm}
\textit{Insights about parameters}

In sensitivity analysis, more variations improve accuracy, but real HPC applications, unlike synthetic evaluations, are resource-intensive. We set a random baseline and incorporate five individual variations per parameter, which are suggested by experts aiming to potentially maximize performance based on that parameter. The number of optimal variations is tailored to factors like dimensionality, runtime distribution concerning each parameter, and computing budget. Concerning the sensitivity analysis targeting the overall runtime, we conducted 100 different \textit{valid} application evaluations in total for each case study. Regarding Case Study 1, $nstb$ is the most influential parameter at 21.71\%, followed by \textit{nkpb} (5\%), \textit{nbatches} (2.46\%), \textit{nstreams} (2.11\%), \textit{tb\_sm\_pair} (2.03\%), \textit{tb\_sm\_zcopy} (1.94\%), \textit{u\_vec} (1.73\%), and \textit{tb\_vec} (1.39 \%). With respect to Case Study 2, \textit{nkpb} comprises 61\% of variability, followed by \textit{nstb} (39\%), \textit{nbatches} (16\%), \textit{nstreams} (2\%), and \textit{ uvec, udscal, tbpair} ($1.5\%$). The presence of several k-points in Case Study 2 emphasizes the significance of \textit{nkpb} in the tuning. Some residual variabilities may be related to runtime uncertainty in HPC applications. Also, the FFT sizes are different between Case Studies, leading to dissimilar importance of the parameters. The evaluated samples show significant runtime variability of up to one order of magnitude.

 Expanding our dataset with additional 100 evaluations for each Case Study, we conduct feature importance and Pearson analyses. Regarding feature importance for modeling, \textit{nstb} at 79.5\%, \textit{nkpb} at 5.1\%, and \textit{tb\_dscal} at 1.5\% dominate for Case Study 1; while  \textit{nkpb} (47\%), \textit{nbatches} (13\%), \textit{nstb} (13\%), \textit{tb\_dscal} (3.7\%), and \textit{u\_vec} (3\%) drive Case Study 2. The Pearson correlation analysis showcases that threadblock size and active threadblocks per SM exhibit around 0.6 correlation due to the maximum number of active threads allowed per SM, suggesting grouping them on the same search. Similar results are obtained for Case Study 2.\\


\vspace{-3mm}
\textit{Inferring independent routines}
  
In addition to the global runtime sensitivity analysis conducted previously, we require a separately sensitivity analysis for each \textit{group} in RT-TDDFT to infer interdependence between them. Table \ref{sensitivity} and \ref{sensitivity2} illustrate the variability results for the three groups and its outer region, Slater Determinant, on Case Study 1 and 2, respectively. The Slater Determinant component is substantially impacted by \textit{nstb}. This parameter determines the number of locally computed bands, which defines the loop iterations, with parameters \textit{nbatches} and \textit{nstreams} affecting how these iterations overlap. Furthermore, \textit{nbatches} links to Groups 1, 2, and 3 due to its impact on memory-transaction efficiency and workload distribution among the kernels. Group 1 (which encompasses the \texttt{3D-cuFFT}, \texttt{cuVec2Zvec}, and \texttt{cuZcopy} kernels) is not significantly influenced by other external parameters. Group 2 (involving the \texttt{cuPairwise} kernel) is not affected by Group 1 parameters; thus, we can infer a weak interdependence. However, Group 3 (consisting of \texttt{3D-cuFFT}, \texttt{cuZvec2Vec}, \texttt{cuDscal}, and \texttt{cuZcopy} kernels) is unexpectedly affected by Group 2's  \textit{tb\_PAIR}/\textit{tb\_sm\_PAIR} (correlated) parameters, which could be attributed to GPU-cache effects, demonstrating interdependence between these two groups. \\

\begin{table}[t!]
\centering
\resizebox{9cm}{!}{%
\begin{tabular}{ll|ll|ll|ll}
\hline
\multicolumn{2}{c|}{\textbf{Group 1}} & \multicolumn{2}{c|}{\textbf{Group 2}} & \multicolumn{2}{c|}{\textbf{Group 3}} & \multicolumn{2}{c}{\textbf{Slater Deter.}} \\
\cmidrule(lr){1-2} \cmidrule(lr){3-4} \cmidrule(lr){5-6} \cmidrule(lr){7-8}
\textbf{Feature} & \textbf{Variability} & \textbf{Feature} & \textbf{Variability} & \textbf{Feature} & \textbf{Variability } & \textbf{Feature} & \textbf{Variability } \\
\midrule
nbatches & 357.33\% & nbatches & 320.62\% & nbatches & 94.81\% & nstb & 88.42\% \\
uvec & 2.96\% & tbsmvec & 3.61\% & tbsmpair  &76.46\%& nbatches & 45.66\% \\
uzcopy   &  1.37\%  & upair & 3.61\% & tbzcopy   &  38.77\% &  nstreams & 39.40\% \\
tbzcopy   &0.99\% & tbpair & 1.03\% & tbdscal & 24.94\%  &  tbdscal    &              6.49 \% \\
tbsmdscal & 0.84\% & uvec& 0.69\% & udscal & 14.26\% &  tbsmpair  &              6.18\% \\
tbvec & 0.68\% & tbsmdscal & 0.69\% & nstreams & 14.14\% & nkpb        &              4.47\% \\
tbdscal & 0.68\% & tbvec & 0.69\% &  uzcopy & 12.96\% & tbsmvec   &              3.95 \%\\  
tbsmvec & 0.68\% & stream & 0.44\% & tbsmzcopy & 9.33\% &tbsmzcopy &              3.92 \%\\
nkpb & 0.38\%      & nstb & 0.34\% & tbsmdscal & 9.31\% &  uvec     &              3.61 \% \\
nstreams & 0.27\% & tbdscal & 0.00\% & uzvec & 9.08\% &  tbvec     &              3.57\% \\
\hline
\end{tabular}%
}
\captionsetup{font=small}
\caption{Sensitivity Analysis: Top 10 sensitive parameters for different routines' runtimes on Case Study 1.}
\label{sensitivity}
\end{table}

\begin{table}[t!]
\centering
\resizebox{9cm}{!}{%
\begin{tabular}{ll|ll|ll|ll}
\hline
\multicolumn{2}{c|}{\textbf{Group 1}} & \multicolumn{2}{c|}{\textbf{Group 2}} & \multicolumn{2}{c|}{\textbf{Group 3}} & \multicolumn{2}{c}{\textbf{Slater Deter.}} \\
\cmidrule(lr){1-2} \cmidrule(lr){3-4} \cmidrule(lr){5-6} \cmidrule(lr){7-8}
\textbf{Feature} & \textbf{Variability} & \textbf{Feature} & \textbf{Variability} & \textbf{Feature} & \textbf{Variability } & \textbf{Feature} & \textbf{Variability } \\
\midrule
nbatches  & 323\% & nbatches & 356\% & nbatches & 49\% & nbatches & 73\% \\
tbvec  & 11\% & nkpb & 8.33\% & tbzcopy & 47.39\% & nstb & 55.4\% \\
tbsmzvec  & 10.48\% & upair & 8.33\% & tbdscal & 47.39\% & tbsmpair & 17.9\% \\
nstreams   & 7.06\% & tbsmpair & 8.33\% & tbpair & 25.64\% & udscal & 17.9\% \\
tbdscal    & 6.45\% & tbsmzcopy & 8.33\% & uzvec & 19.46\% & tbsmvec & 15.77\% \\
uvec       & 5.38\% & tbsmzvec & 8.33\% & nstreams & 16.26\% & tbsmzcopy & 15.12\% \\
tbpair    & 4.84\% & tbzcopy & 8.33\% & tbsmzcopy & 15.69\% & nstreams & 12.8\% \\
tbsmvec     & 4.84\% & uvec & 8.33\% & tbsmdscal & 12.31\% & uvec & 12.5\% \\
tbzcopy   & 4.84\% & uzcopy & 8.33\% & uzcopy & 8.24\% & tbvec & 11.9\% \\
upair     & 4.44\% & tbsmdscal & 6.25\% & tbvec & 5.66\% & uzvec & 11.4\% \\
\hline
\end{tabular}%
}
\captionsetup{font=small}
\caption{Sensitivity Analysis: Top 10 sensitive parameters for different routines' runtimes on Case Study 2.}
\label{sensitivity2}
 \vspace{-4mm}
\end{table}

\vspace{-3mm}
\textit{ Establishing the set of tuning searches. }

\begin{figure}[t!]
\centering 
\includegraphics[scale=0.25]{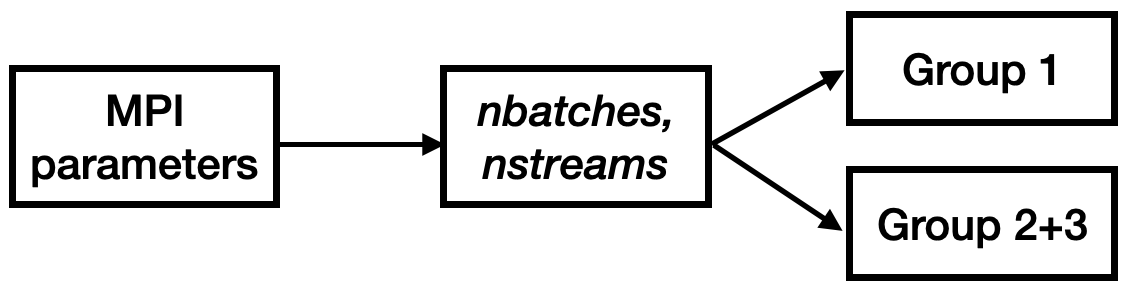}
\captionsetup{font=small}
\caption{Diagram of the resulting dependencies on searches}
\label{search_schema}
 \vspace{-4mm}
\end{figure}

 In our scenario, results for Case Study 1 and Case Study 2 yielded similar conclusions; therefore, the same search strategy is executed for both material systems. We have applied a 10\% cut-off strategy for pruning relationships of parameters linking to other groups. Based on this, MPI parameters do not intersect with the individual execution of the GPU kernels. However, the batch size (\textit{nbatches}), a parameter dictating the number of bands computed in a single invocation and thus impacting the workload per kernel, affects the three groups of GPU routines. Consequently, it lacks orthogonality to them. As such, owing to the necessity of maintaining a uniform batch value across kernels, we first determine the batch value that optimizes the overall execution of the Slater Determinant region. Similarly, MPI parameters do not need to consider the optimal value of \textit{nstreams} and \textit{nbatches} parameters, but rather, they need to align with the number of Slater Determinant iterations to be computed, which depends on the MPI \textit{nstb} parameter. Lastly, parameters belonging to Group 1 exhibit weak variability (under 10\%) to those in Groups 2 and 3. Nevertheless, data reveals an unexpected influence of Group 2 parameters over Group 3 (greater than 10\%), necessitating their joint consideration during the search process. Figure \ref{search_schema} shows a diagram with these dependencies.

Both Group 1 and Group 3 include calls to the same \texttt{cuZcopy} kernel, and  the current application implementation restrics the use of identical parameter values for a kernel across all instances. As per the provided methodology guidelines, our focus was on optimizing the kernel within the region with highest impact, Group 3. Consequently, Group 1's optimization only includes \texttt{cuVec2Zvec} parameters. In the combined search of Group 2+3, the total count of GPU parameters reaches 12 (\texttt{cuPairwise}, \texttt{cuZvec2Vec}, \texttt{cuDscal}, and \texttt{cuZcopy} kernels), surpassing the 10-per-search limit stipulated in our guideline. Leveraging insights from sensitivity analysis and feature importance analysis, we exclude \textit{tb\_zvec} and \textit{tb\_sm\_zvec}, which are assigned values of 64 and 1, respectively. It should be noted that the only tuning parameters impacting the \texttt{cuFFT} routine are \textit{nbatches} and \textit{nstreams}. Table \ref{lower} shows the ultimate set of searches suggested by our methodology.

\begingroup
\linespread{0.75} 
\begin{table}[t!]
\small
\centering
\begin{tabular}{p{1.5cm} p{0.35cm} p{5.3cm}}
\hline
\textbf{Search} & \textbf{Dim} & \textbf{Parameters} \\
\hline
MPI Grid & 3 & \texttt{nstb, nkpb, nspb} \\
Iterations & 2 & \texttt{nbatches, nstreams} \\
Group 1 & 3 & \texttt{u\_VEC, tb\_sm\_VEC, tb\_VEC} \\
Group 2+3 & 10 &  \texttt{u\_PAIR, tb\_sm\_PAIR, tb\_PAIR}\\&& \texttt{u\_ZCOPY,tb\_ZCOPY,tb\_sm\_ZCOPY}\\ && \texttt{u\_DSCAL,tb\_DSCAL, tb\_sm\_DSCAL, u\_ZVEC} \\
\hline
\end{tabular}
\captionsetup{font=small}
\caption{Lower-dimensional searches generated.}
\label{lower}
\vspace{-2mm}
\end{table}
\endgroup


\begin{figure}[t!]
    \centering
    \begin{subfigure}{0.22\textwidth}
        \includegraphics[width=\linewidth]{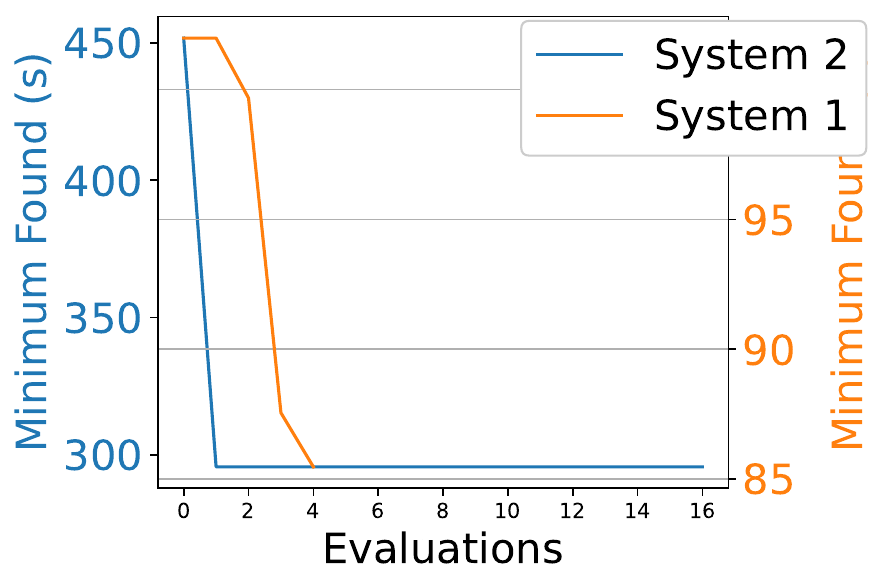}
        \captionsetup{font=small}
        \caption{MPI grid search}
        \label{fig:subfig1}
    \end{subfigure}\hfill
    \begin{subfigure}{0.22\textwidth}
        \includegraphics[width=\linewidth]{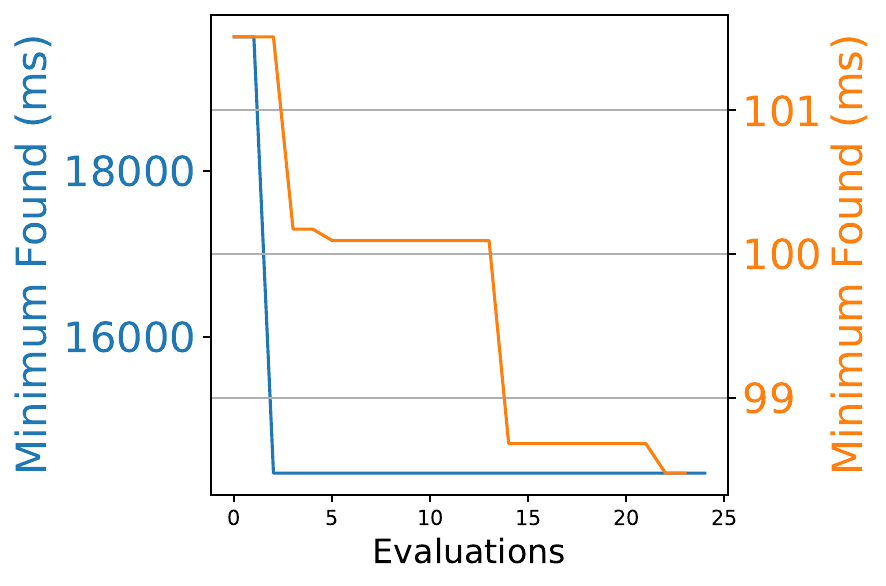}
        \captionsetup{font=small}
        \caption{\textit{batches}+\textit{streams} search}
        \label{fig:subfig2}
    \end{subfigure}
    
    \begin{subfigure}{0.22\textwidth}
        \includegraphics[width=\linewidth]{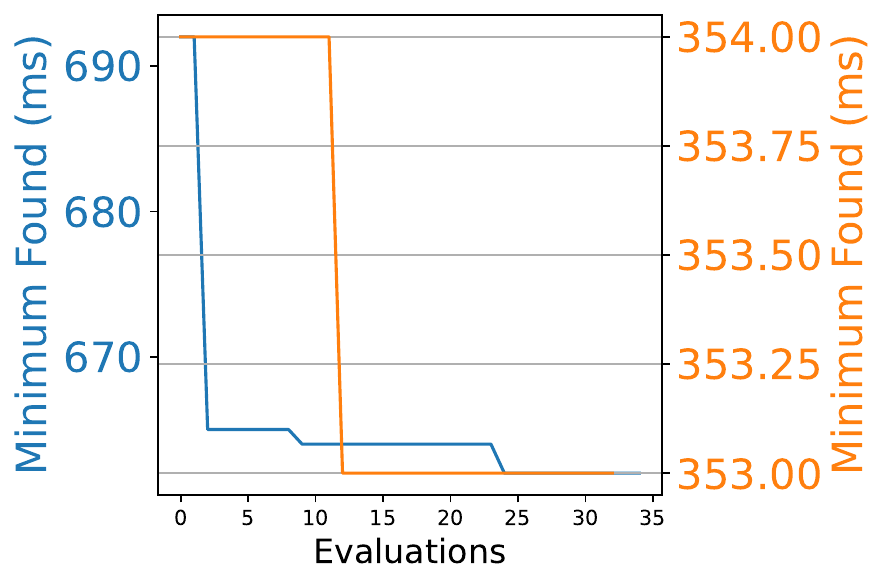}
        \captionsetup{font=small}
        \caption{Group 1 search}
        \label{fig:subfig3}
    \end{subfigure}\hfill
    \begin{subfigure}{0.22\textwidth}
        \includegraphics[width=\linewidth]{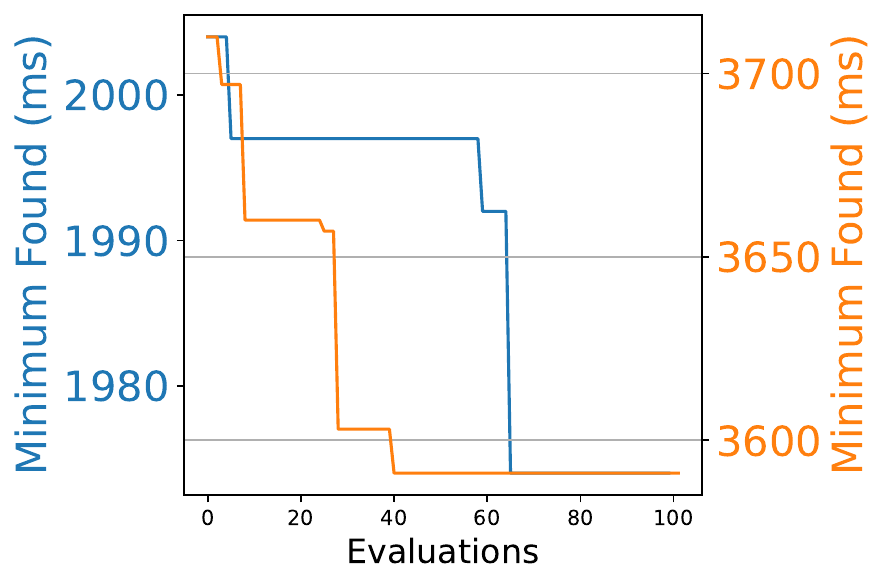}
        \captionsetup{font=small}
        \caption{Group 2+3 search}
        \label{fig:subfig4}
    \end{subfigure}
    \captionsetup{font=small}
    \caption{Progression of the optimal configuration identified by the BO searches over the number of evaluated candidates.}
    \label{fig:four_subfigures}
     \vspace{-5mm}
\end{figure}

Finally, Figure \ref{fig:four_subfigures} depicts the evolution of the optimal configuration found by each \texttt{BO} search over the number of evaluated candidates. In the specific context of \textit{Case Study 1} (orange line), the system has a single k-point and spin. Therefore no parallelization can be exploited over these dimensions, and we set $nspb=1$ and $nkpb=1$, effectively reducing the MPI-grid partition exploration. Additionally, since the number of electron bands is equal to 64, only divisors of this value are tested for the $nstb$ MPI dimension to ensure work balance among ranks. This search constraint, which application and platform experts can provide, saves search time compared to tools that focus solely on tuning unconstrained search spaces. Consequently, the narrowed set of final possibilities, which has already been explored in the sensitivity analysis, allows obtaining the MPI-grid optimal partition without incurring the overhead of a guided \texttt{BO} search. It should be observed that the depicted objective runtimes in Figure \ref{fig:four_subfigures} corresponds to the execution time of the corresponding routine being tuned. Similarly, results for Case Study 2 are depicted with a blue line. To improve the search results, we have used transfer learning to benefit from Case Study 1's configuration database and increase the accuracy of the optimization search exploring space regions that led to good minima in Case Study 1. Transfer-learning improvements on tuning searches have been analyzed in different studies \cite{PMBS,GPTuneCrowd}. As Case Study 2 has 36 k-points and 64 bands, we have constrained the MPI grid search to only multiples of these numbers' divisors to avoid work unbalance or idle MPI ranks. Additionally, we have limited the execution of suggested configurations with a timeout of 15 minutes to reduce the search time.

Conducting the joint 20-dimension search of our Case Studies proved unfeasible to suggest candidates using GPTune due to the required constraints on the expansive search space. We found the same issue with a 17-dimensional search when only targeting the GPU parameters.  This underscores the critical role of our methodology in going beyond the limitations of existing state-of-the-art Bayesian optimization frameworks. On the other hand, the joint Group 2+3 strategy suggested by our methodology outperforms the strategy of independent searches for Group 2 and 3 with a 1\% improvement in Case Study 1 (averaging 5 application executions). It is noteworthy that opting for an independent search for Group 3 eliminates the need to discard any performance parameter, precisely amounting to 10 parameters. Our methodology successfully identified a weak interdependence between Group 2 and 3, and we adhered to a strict 10\% cut-off.  The flexibility of our methodology permits the selection of different cut-offs based on user requirements. In our scenario, there was potential to raise this cut-off slightly to neglect the interdependence and conduct separate searches, as the observed performance gains, though existing, were not remarkably substantial.  However, conducting two independent searches of \textit{N=30} and \textit{N=100} evaluations consumes more resources than the single joint Group 2+3 search of \textit{N=100}, which also obtained better performance. The achieved modest performance gains exhibit noteworthy scalability benefits for larger material systems, leading to cumulative computing savings across multiple executions of the application. In Case Study 2, the joint Group 2+3 search similarly realized a performance improvement of 4.6\% compared to separate Group 2 and Group 3 searches. These results underscore the effectiveness of our methodology in tuning a real HPC application, showcasing its success in challenging scenarios where typical extreme cases are less effective, or even not feasible.

\section{Conclusions}
\label{conclusions}

Our work focus on the critical realm of complex tuning searches within the domain of HPC. Traditional optimization methods, which rely on empirical searches or analytical models, fall short in facing high dimensionality due to vast exploration spaces and intricate interdependencies among parameters and routines to tune. Finding these interdependencies may incur a significant cost with no guarantees to succeed. 
As a consequence, many programmers limit the parameters subjected to tuning, missing out on the full potential of modern supercomputers. They often resort to separate tuning searches for different routines, neglecting the interplay among them. Alternatively, they conduct a single joint search for all routines, a costly and inaccurate approach as search mechanisms struggle with high dimensionality. To address this challenge, we propose a methodology that navigate complex tuning search spaces characterized by a high number of parameters and different levels of interdependence.  We demonstrate the effectiveness of our methodology on Real-Time Time-Dependent DFT applications.

Our methodology follows five premises to efficiently find an optimized set of searches for tuning several routines within an application. Through a bottom-up design composed of two phases, the methodology suggests an optimized set of tuning searches that can be handled by \texttt{BO}, thereby increasing solution quality while optimizing computational resources. In the first phase, it tags the influence of different tuning parameters on each routine with an influence score. To determine this influence, we conduct an analysis based on runtime sensitivity to individual parameter variations, which demonstrates to capture interdependence while significantly reducing the required observations compared to the orthogonal analysis proposed in literature. In the second phase, our approach creates a Directed Acyclic Graph derived from the obtained scores, which represents the interplay between routines and performance parameters, and solve its partition based on a defined interdependence cut-off: Routines that are linked to others by external parameters must be explored together. Notably, an extremely low cut-off resulting in a merged search of higher dimensionality may not compensate when employing Bayesian optimization with minimal evaluations. The wider search space generated may become challenging to navigate in a limited number of evaluations. Additionally, the methodology constrains each resulting search to a maximum of 10 dimensions, selectively discarding less crucial parameters when necessary. This restriction allows \texttt{BO} to navigate the search space more efficiently with minimal evaluations, a preference in HPC tuning.

Exemplified through five synthetic functions and a RT-TDDFT application, our methodology succeeded in finding effective configurations in search spaces of 20 parameters with different levels of interdependence. Specifically, compared to the typical extremes of fully-independent searches and a single joint search, our methodology on the synthetic functions identified configurations resulting in quality improvements of up to 8\%, reducing the required search time by up to 95\%. While the GPU-offloaded RT-TDDFT version creates a constrained tuning search of 20 parameters that cannot be directly explored as a single joint search with state-of-the-art HPC autotuners, our methodology successfully addresses and resolves this challenge, demonstrating its effectiveness across different tuning scenarios. Our work not only reduces the required number of evaluations to conduct the search, a costly aspect in HPC, but also contributes to bridging the gap in the lack of a dedicated BO-based tuning approach for a large set of tuning parameters in HPC.

\bibliographystyle{IEEEtran}
\bibliography{latex/mybibfile}

\end{document}